\documentclass[prl,aps,twocolumn,showpacs, floatfix]{revtex4}
\usepackage{amssymb}
\usepackage{graphicx}
\usepackage{dcolumn}
\usepackage{bm,amsmath,verbatim}

\def\be{\begin{equation}}
\def\ee{\end{equation}}
\def\bea{\begin{eqnarray}}
\def\eea{\end{eqnarray}}
\def\bse{\begin{subequations}}
\def\ese{\end{subequations}}

\def\be{\begin{eqnarray}}
\def\ee{\end{eqnarray}}

\begin{document}

\title{Hole-doped semiconductor nanowire on top of an \textit{s}-wave
superconductor: A new and experimentally accessible system for Majorana
fermions}
\author{Li Mao$^{1}$}
\author{Ming Gong$^{1}$}
\author{E. Dumitrescu$^{2}$}
\author{Sumanta Tewari$^{2}$}
\author{Chuanwei Zhang$^{1}$}
\thanks{cwzhang@wsu.edu}

\begin{abstract}
Majorana fermions were envisioned by E. Majorana in 1935 to describe
neutrinos. %So far most
%candidate solid-state systems for Majorana fermions suffer from one important problem:
%the required experimental parameters are beyond the capacity of current
%experiments.
Recently it has been shown that they can be realized even in a class of
electron-doped semiconductors, on which ordinary $s$-wave superconductivity
is proximity induced, provided the time reversal symmetry is broken by an
external Zeeman field above a threshold. Here we show that in a hole-doped
semiconductor nanowire the threshold Zeeman field for Majorana fermions can
be very small for some \emph{magic} values of the hole density. In contrast
to the electron-doped systems, smaller Zeeman fields and much stronger
spin-orbit coupling and effective mass of holes %coupled with other
%topical advantages of holes in contrast to electrons,
allow the hole-doped systems to support Majorana fermions in a parameter
regime which is routinely realized in current experiments.
%A hole-doped nanowire, with its fundamentally different underlying physics
%from its electron-doped counterpart, also leads to many other topical
%advantages for realizing a topologically non-trivial state. Thus, this
%system is uniquely suited among all solid state systems for realizing and
%manipulating Majorana fermions in controllable experiments.

\begin{comment}

The emergence of Majorana fermions in solid state systems is not only an
extraordinary phenomenon in itself, but also has sparked tremendous recent
attention for their potential use in fault tolerant quantum computation \cite%
{Kitaev,Nayak-RMP}. It has been recently shown \cite{Sau, Alicea,Long-PRB,
Roman, Oreg, Lutchyn} that electron-doped semiconductor thin films and
wires, in proximity contact with \textit{s}-wave superconductors, support
Majorana excitations. Experimental observation of Majorana fermions in
electron-doped systems can still be challenging because of the stringent
requirements of large magnetic fields and a low carrier density. Here we
show that a one-dimensional \emph{hole-doped} semiconductor nanowire is
essentially free of these problems and uniquely suited among all solid state
systems for realizing Majorana fermions. We show that, for some
\textquotedblleft magic" values of the chemical potential, it can support
Majorana fermions for \emph{vanishingly small} values of the Zeeman field.
This, coupled with a higher effective mass and a larger spin-orbit coupling
of holes, leads quite remarkably to a required parameter regime which is
already attainable in current experiments.

\end{comment}
\end{abstract}

\affiliation{$^{1}$Department of Physics and Astronomy, Washington State University,
Pullman, WA 99164 USA \\
$^{2}$Department of Physics and Astronomy, Clemson University, Clemson, SC
29634 USA}
\maketitle

Recently some exotic condensed matter systems, such as the Pfaffian states
in fractional quantum Hall (FQH) systems~\cite%
{Moore,Nayak-Wilczek,Read,dassarma_prl'05}, chiral $p$-wave
superconductors/superfluids~\cite%
{Ivanov,DasSarma_PRB'06,Volovik,Volovik2,tewari_prl'2007}, topological
insulator (TI)~\cite{fu_prl'08}, as well as a ferromagnet-superconductor
heterostructure \cite{Lee2}, have been proposed as systems supporting
quasiparticles with non-Abelian statistics \cite{Nayak-RMP}. These systems
allow a special type of quasiparticles called Majorana fermions which
involve no energy cost.
%Majorana fermions were\ first envisioned by E. Majorana in 1935 to
%describe neutrinos.
The second quantized operators $\gamma _{i}$ for the Majorana excitations
are self-hermitian, $\gamma _{i}^{\dagger }=\gamma _{i}$ (particles are
their own anti-particles), which lies at the heart of their non-Abelian
statistical properties. %The self-hermitian property
%lies at the heart of the non-Abelian anyonic statistics \cite%
%{Nayak-Wilczek,Read,Ivanov} of these excitations.
Due to the fundamental difference of Majorana fermions from any other known
quantum particles in nature, the emergence of these particles in solid state
systems would in itself be an extraordinary phenomenon. Their potential use
in fault-tolerant topological quantum computation (TQC) \cite{Nayak-RMP}
makes their realization in controllable solid state systems even more
significant. %from a
%long term technological point of view.

It has been shown recently \cite{Sau,Alicea,Long-PRB,Roman,Oreg,Lutchyn}
that an electron-doped semiconducting thin film or nanowire with a sizable
spin-orbit coupling can host, under suitable conditions, Majorana fermion
excitations localized near defects. This proposal followed on an earlier
similar proposal in the context of cold atomic systems \cite{Zhang}. When
the film or the nanowire is in the presence of a Zeeman splitting $V_{z}$
(with Land\'{e} factor $g_{e}^{\ast }$) and an $s$-wave superconducting pair
potential $\Delta $, which can be proximity induced by a nearby
superconductor, the system enters into a topological superconducting (TS)
state for
\begin{equation}
V_{z}^{2}>\Delta ^{2}+\mu ^{2},\quad V_{z}=g_{e}^{\ast }\mu _{B}B/2
\label{Eq:Parameters}
\end{equation}%
Here $\mu $ is the chemical potential in the semiconductor which is
controlled by the density of doped electrons. %Solutions of the Bogoliubov
%de-Gennes (BdG) equations and/or topological arguments have been employed to
%show that in the TS state on a nanowire the two end points of the wire
%support two zero energy Majorana fermion excitations.
Despite the theoretical success, the requirement Eq.~(\ref{Eq:Parameters})
for the TS state in an electron-doped nanowire leads to two obvious
experimental challenges: a low electron density and a high magnetic field.
For a small carrier density a nanowire tends to become insulating due to the
strong disorder-induced fluctuations of the chemical potential. A high
magnetic field, on the other hand, can be detrimental to pairing as well as $%
s$-wave proximity effect itself.
\begin{comment}
The Zeeman field $V_{z}$, which is
usually applied by a magnetic field external to the nanowire, must be larger than the
chemical potential $\mu$ and the pairing potential $\Delta $. For a typical value
of Zeeman field $V_{z}\sim 2$ meV (corresponds to $g_{e}^{\ast }\sim 35$ and
$B\sim $ 2 Tesla for an InAs nanowire), this means that the carrier
density $n$ in the wire must be low ($n\sim 10^{4}$ cm$^{-1}$).
In
 electron-doped nanowires such a small carrier density is experimentally
unattainable. This is because the system becomes insulating due to
disorder-induced chemical potential fluctuations for such small values of
the carrier density.
%For larger values of the carrier densities, the
%threshold $V_{z}$ can be very high, making it detrimental for the adjacent
%proximity-inducing superconductor.
By including more than one
confinement-induced bands in the nanowire, it has been recently shown that \cite{Lutchyn}
the allowed values of $n$ in the TS state can be brought to a range of $%
n\sim 10^{5}$ cm$^{-1}$ for $B\sim 4$ Tesla in an InAs nanowire. (For a similar idea for
chiral-$p$ wave superconductors see Ref.~[\onlinecite{Potter}].) Although an
improvement, this is still not in the experimentally accessible regime ($%
n\sim 10^{6}$ cm$^{-1}$). Furthermore, such a high magnetic field $B\sim 4$ Tesla can be
detrimental to the pairing in the nearby superconductor as well as the proximity effect itself.
\end{comment}

In this Letter we show that a \emph{hole-doped} semiconductor nanowire can
solve all these problems encountered in the electron-doped systems. The
hole-doped nanowire is very different in many respects from its
electron-doped counterpart due to its different band structure and the value
of the effective spin of the carriers. For some \textquotedblleft magic"
values of the carrier (hole) density the threshold Zeeman splitting for the
TS states and Majorana fermions can become very small, therefore the
constraint on the carrier density as given in Eq.~(\ref{Eq:Parameters}) is
absent for the hole-doped nanowires.
%Furthermore, in the conduction band of the
%electron-doped semiconductors the electrons have a relatively small
%effective mass and spin-orbit coupling. This also is responsible for a low
%carrier density in the TS state in the electron-doped systems.
%semiconductors physically originates from the small effective mass and
%spin-orbit coupling of the electrons in the conduction band of the
%semiconductor \cite{Vurgaftmana}. the larger effective mass
Furthermore, the effective mass and spin-orbit coupling in the \emph{p}-type
\emph{valence} band holes are much larger than electrons, which leads to a
larger Fermi vector $k_{F}$. This larger $k_{F}$ leads to a larger required
carrier density ($\sim 10^{6}$ cm$^{-1}$) for the TS state, which,
remarkably, is now \emph{routinely achieved} in many experiments \cite%
{ptype,GaSb,Lieber}. The large carrier density provides strong screening of
the disorder potentials, leading to much smaller fluctuations of the
chemical potential \cite{disorder} in the nanowire. Furthermore, the small
ratio between the Zeeman coupling and the spin-orbit energy (orders of
magnitude smaller than that in the electron-doped systems) leads to a small
carrier mobility requirement for the hole-doped TS state (3 order of
magnitude smaller than that for the electron-doped TS state), as pointed out
recently in \cite{sau3}. Let us also point out that the superconducting
proximity effect on a hole-doped nanowire has been observed in recent
experiments \cite{Lieber}.
%As such, the only new advance needed to induce a
%TS state in such a nanowire is applying an external Zeeman splitting, which
%can be relatively small in magnitude (compared with an electron-doped
%system) near the magic values of the carrier density.
It seems therefore that a Majorana-carrying TS state is tantalizingly close
to experimental reach in a hole-doped nanowire.
\begin{comment}
Apart from the immense
experimental implications, this work also has a much broader fundamental
significance because of the possibility of semiconductor Majorana fermions
with a minimal TRS breaking, which requires further exploration.
\end{comment}%
\begin{comment}
There are other important advantages of using hole-doped
systems as well. Due to the requirement of a small $V_{z}$, the topological
state can be much more robustly realized avoiding the detrimental effects of
the magnetic field on the nearby superconductor as well as on the $s$-wave
proximity effect itself. The small Zeeman splitting also permits the
possibility of observing Majorana fermions with a wide range of
semiconductor materials which may not have a large $g$ factor \cite%
{GaSb,Lieber}.
\end{comment}

\begin{figure}[t]
\centering
\includegraphics[width=1\linewidth]{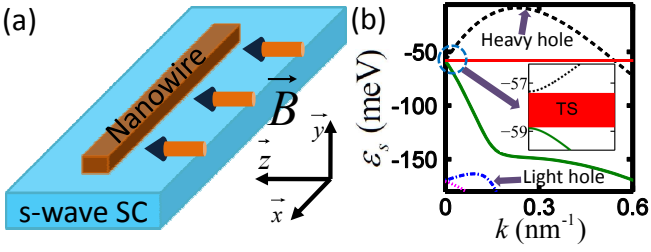}
\caption{(a) Schematic plot of the experimental setup. (b) Illustration of
the valence band structure of 1D hole-doped nanowire with a Zeeman field.
Only the lowest band along the $y$ and $z$ directions is considered. There
are two heavy hole (dashed black and solid green curves) and two light hole
(blue dash-dotted and pink dotted curves) bands. The thick red line
(amplified as shadow in inset) gives the regime of chemical potential with a
single Fermi surface, which leads to the topological superconducting state.
The parameters are for a hole-doped InAs nanowire with $\protect\gamma %
_{1}=20$, $\protect\gamma _{2}=8.5$. $\protect\alpha =3.3\times 10^{5}$ m/s,
$V_{z}=1.5$ meV, $L_{z}=14$ nm, $L_{y}=10$ nm.}
\label{setup}
\end{figure}

\emph{Set-up and Hamiltonian:} The experimental setup is illustrated in Fig. %
\ref{setup}a, where a hole-doped semiconductor nanowire is placed on top of
an $s$-wave superconductor. We focus on large-band gap semiconductors, where
the single particle Hamiltonian of holes is described by the four-band
Luttinger model \cite{SCZhang} (henceforth we set $\hbar =1$, and$\ m=-1$),
\begin{equation}
H_{L}=({\frac{\gamma _{1}}{2}}+{\frac{5\gamma _{2}}{4}})\nabla ^{2}-\gamma
_{2}(\nabla \cdot \mathbf{J})^{2}-i\alpha (\mathbf{J}\times \nabla )\cdot
\hat{z}+V_{z}J_{z}-\mu ,  \label{eq-HL}
\end{equation}%
where $\alpha $ is the Rashba spin-orbit coupling due to the
inversion-symmetry breaking, and the fourth term is the Zeeman field $%
V_{z}=g_{h}^{\ast }\mu _{B}B$ generated by the external magnetic field along
the $z$ direction. $\mathbf{J}$ is the total angular momentum operator for a
spin-3/2 hole, $\gamma _{1}$ and $\gamma _{2}$ are the Luttinger parameters.
Note that here the relation between $V_{z}$ and $B$ differs from Eq. (\ref%
{Eq:Parameters}) by a factor 1/2 because we use the spin-3/2 matrix $J_{z}$,
instead of the Pauli matrix which was used for Eq. (\ref{Eq:Parameters})
\cite{Sau} (thus the required $B$ for the same $V_{z}$ is smaller by a
factor 1/2).

To simplify the calculations we assume a rectangular cross-section of the
nanowire with the widths $L_{y}$ and $L_{z}$. The strong confinement along
the $y$ and $z$ directions makes the energy levels quantized on these axes.
To illustrate the emergence of the Majorana fermions, we first consider a
single band, \textit{i.e.}, the lowest energy state along the $y$ and $z$
directions. Using the ground state wavefunction $\phi (y,z)$ along the $y$
and $z$ directions, the original Hamiltonian (\ref{eq-HL}) in 3D can be
projected to an effective 1D form%
\begin{eqnarray}
H_{1}(x) &=&\left( \gamma _{1}/2+5\gamma _{2}/4-\gamma _{2}J_{x}^{2}\right)
\partial _{x}^{2}+\pi ^{2}\gamma _{2}J_{y}^{2}/L_{y}^{2}  \notag \\
&&+\pi ^{2}\gamma _{2}J_{z}^{2}/L_{z}^{2}+i\alpha J_{y}\partial
_{x}+V_{z}J_{z}-\xi -\bar{\mu},  \label{1DHam}
\end{eqnarray}%
where $\xi =5\gamma _{2}\pi ^{2}\left( L_{y}^{-2}+L_{z}^{-2}\right) /4$, and
$\bar{\mu}=\mu +\gamma _{1}\pi ^{2}\left( L_{y}^{-2}+L_{z}^{-2}\right) /2$
is the shifted chemical potential due to the confinement.

In Fig. \ref{setup}b, we plot the energy spectrum $\varepsilon _{S}$ of the
Hamiltonian (\ref{1DHam}) for holes in an 1D geometry. There are two heavy
hole and two light hole bands.
%We are interested in the chemical potential regime around the
%heavy hole bands.
If $\mu $ lies in the shaded region, it intercepts only one Fermi surface.
An odd number of Fermi surfaces implies a breakdown of the fermion doubling
theorem (due, in this case, to the Zeeman splitting), which yields, in the
presence of a superconducting pair potential, the required TS state for
Majorana fermions \cite{fu_prl'08}. The position of the shaded region (the
TS state) is determined by the Hamiltonian (\ref{1DHam}) at $k=0$ that
depends on the parameters $\gamma _{1}$, $\gamma _{2}$, $L_{y}$, $L_{z}$,
while the width of the shaded region is determined by $V_{z}$. The large
effective mass and strong spin-orbit coupling of the holes lead to a high
density $n=\int_{0}^{k_{F}}dk/2\pi \sim 10^{6}$\ cm$^{-1}$\ of holes in the
TS state. Such a high density is routinely realized in nanowire experiments
\cite{ptype,GaSb,Lieber}. The high carrier density provides a strong
screening of the disorder potential, suppressing the spatial chemical
potential fluctuations and disorder effects \cite{disorder}.

The superconducting pair potential can be induced in the hole-doped nanowire
through the proximity contact with an $s$-wave superconductor (Fig.~\ref%
{setup}a), as demonstrated in experiments \cite{Lieber}. This yields the
Hamiltonian,
\begin{equation}
H_{sc}=\sum\nolimits_{m_{J}}\int d^{3}\mathbf{r}\Delta _{sm_{J}}\left(
\mathbf{r}\right) \hat{\psi}_{m_{J}}^{\dag }\hat{\psi}_{-m_{J}}^{\dag }+%
\text{H.c.,}  \label{eq-Hs}
\end{equation}%
where\ $\hat{\psi}_{m_{J}}^{\dag }$ are the creation operators for holes
with the angular momentum $m_{J}=\frac{1}{2},\frac{3}{2}$ and $\Delta
_{sm_{J}}\left( \mathbf{r}\right) $ is the proximity induced pair potential.
The form of the pairing Hamiltonian is dictated by the fact that $\Delta
_{sm_{J}}$ couples particles with $m_{J}$ with particles with $-m_{J}$, and
should be determined through the microscopic theory of the proximity effect
\cite{unpublished}.

Taking account of the spin-$3/2$ and the particle-hole degrees of freedom in
the superconductor, the Bogoliubov-de-Gennes (BdG) Hamiltonian can be
written as an $8\times 8$ matrix
\begin{equation}
\hat{H}_{\text{BdG}}=%
\begin{pmatrix}
H_{1}(x) & \Delta _{S}(x) \\
\Delta _{S}^{\ast }(x) & -\Upsilon ^{\dag }H_{1}^{\ast }(x)\Upsilon%
\end{pmatrix}
\label{BdG}
\end{equation}%
in the Nambu spinor basis $\hat{\Phi}(x)=\left( \hat{\psi}(x),\Upsilon \hat{%
\psi}^{\dag }(x)\right) ^{T}$ with $\Upsilon =i(I_2\otimes\sigma _{x})\tau
_{y}$ ($I_2$ is the $2\times 2$ unit matrix$),\Delta _{S}(x)=$ diag$\left(
\Delta _{s3/2},\Delta _{s1/2},\Delta _{s1/2},\Delta _{s3/2}\right) $, and $%
\hat{\psi}(x)=(\hat{\psi}_{\frac{3}{2}}(x),\hat{\psi}_{\frac{1}{2}}(x),\hat{%
\psi}_{-\frac{1}{2}}(x),\hat{\psi}_{-\frac{3}{2}}(x))$.

\emph{Parameter space for the topological state: }In a 1D nanowire, the
parameter regime for the Majorana fermions (in 1D the Majorana fermions are
localized at the two end points) can be determined by the topological index $%
\mathcal{M}$ \cite{Kitaev2, Sau2} defined as,
\begin{equation}
\mathcal{M}=\text{sgn}\left[ \text{Pf}\left\{ \Gamma (0)\right\} \right]
\text{sgn}\left[ \text{Pf}\left\{ \Gamma (\pi /a)\right\} \right] .
\label{index}
\end{equation}%
Here Pf represents the Pfaffian of the anti-symmetric matrix $\Gamma
(k)=-iH_{\text{BdG}}(k)(\varsigma _{y}\otimes \Upsilon )$, $\varsigma _{y}$
is the Pauli matrix, $H_{\text{BdG}}(k)$ is the corresponding BdG
Hamiltonian in the momentum space ($-i\partial _{x}\rightarrow k$), and $a$
is the lattice constant. $\mathcal{M}=-1$ ($+1$) corresponds to the
topologically nontrivial (trivial) states with (without) Majorana fermions.
Using the fact that $\Gamma (k)$ is an anti-symmetric matrix that can be
diagonalized by a lower triangular matrix \cite{Bp}, we find
\begin{equation}
\text{Pf}\left\{ \Gamma (k)\right\} =\text{Pf}\left\{ \Delta _{S}\Upsilon
\right\} \text{Pf}\left\{ \Upsilon \Delta _{S}+H_{1}^{T}\left( k\right)
\Upsilon \Delta _{S}^{-1}H_{1}\left( k\right) \right\}  \label{pha}
\end{equation}%
through a straightforward calculation.

In the continuous limit $k=\pi /a\rightarrow \infty $, the $k^{2}$ terms in
the single particle Hamiltonian $H_{1}\left( k\right) $ dominates and all
other terms in $\Gamma (k)$ can be neglected. In this case, it can be shown
that sgn$[$Pf$\left\{ \Gamma (k)\right\} ]=$ sgn$[\det \left( H_{1}\left(
k\right) \right) ]=1$, therefore $\mathcal{M}$ is solely determined by the
sign of Pf$\{\Gamma (0)\}$. Pf\{$\Gamma (0)\}$ can be derived analytically
from Eq. (\ref{pha}), yielding $\mathcal{M}=$ sgn$[\mathcal{F}]$, where
\begin{equation}
\mathcal{F}=f_{0}-f_{1}V_{z}^{2}+9V_{z}^{4}/16,  \label{pha1}
\end{equation}%
$f_{0}=(\bar{\mu}^{2}+\Delta _{s3/2}\Delta _{s1/2}-\beta _{1}^{2}-\beta
_{2}^{2})^{2}+[(\Delta _{s3/2}-\Delta _{s1/2})\bar{\mu}+\beta _{1}(\Delta
_{s3/2}+\Delta _{s1/2})]^{2}$, $f_{1}=\left[ 10\bar{\mu}^{2}+10\beta
_{1}^{2}+16\beta _{1}\bar{\mu}+9\Delta _{s1/2}^{2}+\Delta _{s3/2}^{2}-6\beta
_{2}^{2}\right] /4$, $\beta _{1}=\pi ^{2}\gamma _{2}\left(
L_{z}^{-2}-L_{y}^{-2}/2\right) $, $\beta _{2}=\sqrt{3}\pi ^{2}\gamma
_{2}L_{y}^{-2}/2$. Since $\mathcal{M}$ changes sign when $\mathcal{F}$
changes sign, the phase boundary between the topologically trivial and
nontrivial states can be determined by setting $\mathcal{F}=0$.

\begin{figure}[t]
\centering\includegraphics[width=1\linewidth]{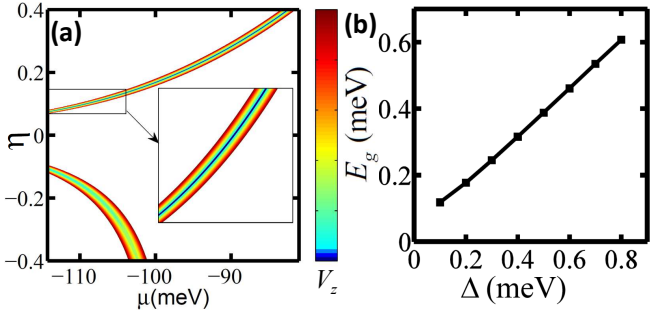}
\caption{(a) Plot of the phase boundary ($\mathcal{F}=0$) between
topological and non-topological states for different $\protect\mu $, $%
V_{z}^{c}$ and $\protect\eta $. $\Delta _{s1/2}=1$ meV, $\protect\beta %
_{2}=55.4$ meV (corresponding to $L_{y}=10$ nm), $\protect\beta _{1}=(1-%
\protect\eta )/(2\protect\sqrt{|\protect\eta |})\protect\sqrt{\protect\beta %
_{2}^{2}-\protect\eta \Delta _{s1/2}^{2}}$. The color region represents the
TS state with $\mathcal{M}$= $-1$. Different colors correspond to different
critical Zeeman fields. (b) Plot of the minigap $E_{g}$ with respect to $%
\Delta $. $\protect\mu =-50.9$ meV, $V_{z}=1.5$ meV, $L_{y}\approx 9$ nm and
$L_{z}\approx 15$ nm. }
\end{figure}

When the ratio $\eta =\Delta _{s3/2}/\Delta _{s1/2}\geq \Delta
_{s3/2}^{2}/\beta _{2}^{2}$, there exists a magic value of the chemical
potential $\bar{\mu}_{0}=\frac{\eta +1}{4\left\vert \eta \right\vert }\sqrt{%
\eta \beta _{2}^{2}-\Delta _{s3/2}^{2}}$\ in the heavy hole band such that $%
f_{0}=0$, where $\beta _{1}=\left( 1-\eta \right) \bar{\mu}_{0}/\left(
1+\eta \right) $. For $f_{0}=0$, $\mathcal{F}<0$\ and Majorana fermions
exist even for a vanishingly small $V_{z}$. Clearly, $\eta \geq \Delta
_{1}^{2}/\beta _{2}^{2}$ requires that $\Delta _{s3/2}$ and $\Delta _{s1/2}$
have the same sign and $\beta _{2}^{2}\geq \Delta _{1}\Delta _{2}$. Because
of the strong confinement, the second condition can be easily satisfied.
Therefore the threshold Zeeman field $V_{z}^{c}$ for the TS state vanishes
%existence of Majorana fermions with a vanishly
%small $V_{z}$ still holds
when $\Delta _{s3/2}$ and $\Delta _{s1/2}$ are of the same sign, independent
of their relative magnitudes. When $\Delta _{s3/2}$ and $\Delta _{s1/2}$
have different signs, $V_{z}^{c}$ becomes nonzero, but is still much smaller
than that for the electron-doped semiconductors. Therefore the relative
signs of the pair potentials should not matter in realistic experiments
because a reasonable $V_{z}$ is always needed to generate a sizable chemical
potential region for the TS state. In Fig. 2, we plot the boundary $\mathcal{%
F}=0$ between topologically trivial and non-trivial states for different $%
\mu $, $V_{z}^{c}$ and $\eta $. The TS states for a fixed $V_{z}^{c}$ are
embraced by two lines with the same color for the corresponding $V_{z}^{c}$.
For instance, for $\eta =0.15$ and $V_{z}^{c}=1$ meV, Fig. 2 shows that the
TS state exists in the region $\mu _{1}<\mu <\mu _{2}$ with $\mu _{1}=-103.3$
meV and $\mu _{2}=-100.8$ meV. Clearly, the $V_{z}=0$ line (the center blue
line) exists for $\eta >0$, but vanishes for $\eta <0$. When $\eta
\rightarrow 0^{+}$, $\bar{\mu}_{0}\rightarrow -\infty $.

The vanishingly small $V_{z}^{c}$ for the TS state at $\eta >0$ may also be
understood by projecting the four band Luttinger model in Eq. (\ref{1DHam})
to an effective two heavy hole band model because of the large energy
splitting between the heavy and light hole bands. The resulting two band
model is generally very complicated for finite $k_{x}$ and $V_{z}$. However,
because the Pfaffian is determined by the Hamiltonian at $k=0$ and we are
interested in the TS state with vanishingly small $V_{z}$, we can do the
band projection at $V_{z}=0$ and around $k_{x}=0$, leading to an effective
pairing $\Delta ^{eff}=\left( \Delta _{s3/2}-\kappa \Delta _{s1/2}\right)
/\kappa $ and an effective chemical potential $\bar{\mu}_{eff}=\bar{\mu}-%
\sqrt{\beta _{1}^{2}+\beta _{2}^{2}}$, with $\kappa =\left( \sqrt{\beta
_{1}^{2}/\beta _{2}^{2}+1}-\beta _{1}/\beta _{2}\right) ^{2}$. Here we have
neglected non-diagonal term of the pairing because $\beta _{2}^{2}\gg \Delta
_{s3/2}\Delta _{s1/2}$. When $\eta >0$, $\Delta ^{eff}$ may vanish by
choosing $\beta _{1}/\beta _{2}=\left( 1-\eta \right) /2\sqrt{\eta }$,
therefore the critical Zeeman field $V_{z}^{c}$ also vanishes when $\bar{\mu}%
_{eff}=0$. While when $\eta <0$, $\Delta ^{eff}$ is always finite and there
is a minimum $V_{z}^{c}$ based on Eq. (\ref{Eq:Parameters}) for the two-band
model. Note that the zero $\Delta ^{eff}$ at $k_{x}=0$ and $V_{z}=0$ does
not imply the zero $\Delta $ at a finite $V_{z}$ and $k_{x}$. For a large $%
k_{x}$, the coupling between heavy and light holes becomes important and the
mini-gap is finite for a large $V_{z}$ even at the magic $\mu _{0}$ (see
Fig. 3b).

Henceforth, we consider two representative cases to further illustrate our
results: (I) $\Delta _{s3/2}=\Delta _{s1/2}=\Delta $\ and (II) $\Delta
_{s3/2}=-\Delta _{s1/2}=\Delta $. For the a reasonable large $V_{z}$, both
pairings yield similar observable signals. The corresponding topological
region is plotted in Fig. 3a. In the case (I), the TS states exist in the
region $|\bar{\mu}-\bar{\mu}_{0}|\lesssim |V_{z}|/2$\ for both positive and
negative $V_{z}$. The transition at $V_{z}=0$\ (at which the superconductor
is gapless and non-topological) is a quantum transition at which neither the
symmetry of the system nor the topological properties change as a function
of $V_{z}$. Note that the system crosses the phase transition boundary lines
twice when $V_{z}$ changes from \ $-\infty $ to $+\infty $ for a fixed $\bar{%
\mu}$, except at the magic value $\bar{\mu}_{0}$. At the phase boundary, the
quasiparticle energy gap closes and the superconductor becomes gapless. At
the magic $\bar{\mu}_{0}$, the two lines of the phase boundary merge at $%
V_{z}=0$, therefore the superconductor is gapped and topological for all $%
V_{z}$ except at $V_{z}=0$.

\begin{figure}[t]
\centering\includegraphics[width=1\linewidth]{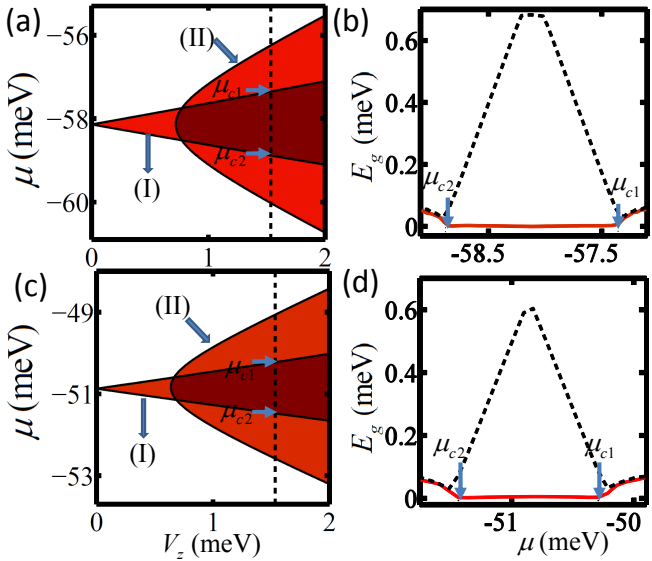}
\caption{The parameter regime for the existence of Majorana fermions. (a,b)
and (c,d) correspond to the single and two band models. $L_{y}\approx 10$ nm
and $L_{z}\approx 14$ nm for (b) and case (I) in (a). $L_{y}\approx 9$ nm
and $L_{z}\approx 15$ nm for (d) and case (I) in (c). $L_{y}\approx 14.2$
nm, $L_{z}\approx 9.5$ nm for case (II) in (a) and (c). The widths of the
nanowire are chosen to obtain a small critical Zeeman field for each case.
(a,c) are obtained from the topological index $\mathcal{M}=-1$ (the filled
regions). (c,d) are obtained from solving the BdG equation with $V_{z}=1.5$
meV. $\Delta =1$ meV. The dashed lines in (b,d) give the minigap. The
parameters $\protect\gamma _{1}$, $\protect\gamma _{2}$, $\protect\alpha $
are the same as that in Fig. 1b. }
\label{phasediag}
\end{figure}

In the case (II), the threshold $V_{z}^{c}\approx p\Delta $ with $p=2\left[
1+2\left( 1+\beta _{2}^{2}/\beta _{1}^{2}\right) ^{-1/2}\right] ^{-1}$ (see
Figs. 3a and 3c). For instance, for hole-doped InAs nanowires (with a
typical $g_{h}^{\ast }=35$ \cite{gfactor}) with Nb as the adjacent
superconductor ($\Delta \simeq 1$ meV), the required magnetic field $B$ is $%
\sim 0.35$ T, which is about 1/3 of the corresponding $B$ for electron-doped
InAs nanowires (with the same material Nb as the adjacent superconductor)
\cite{Roman}, and can be easily realized with a bar magnet without affecting
the superconducting pairing. Therefore, we find that irrespective of the
relative sign of $\Delta _{s3/2}$ and $\Delta _{s1/2}$ (which can only be
determined from a more microscopic calculation \cite{unpublished}) the
threshold $B$ for the TS state in the hole-doped case is much smaller than
that in the electron-doped case. The fact that the Majorana fermions can be
observed even with a small magnetic field opens the possibility of using a
wide range of semiconductor materials with only small $g_{h}^{\ast }$
factors \cite{GaSb,Lieber} and is one big advantage of using hole-doped
semiconductors.

To further confirm the existence of the Majorana fermions in the above
parameter regime, we also numerically solve the BdG equations (\ref{BdG})
and obtain the energy spectrum and eigenstates. The Majorana fermion
corresponds to a zero energy eigenvalue in the BdG spectrum. Henceforth, we
present our results only for the case (I), but have confirmed that the
results for the case (II) are similar. In Fig. \ref{phasediag}b, we plot the
ground and the first excited state energies. The ground state energy becomes
zero in the region $\mu _{c_{2}}<\mu <\mu _{c_{1}}$. In the same parameter
region, the topological index $\mathcal{M}=-1$ (Fig. \ref{phasediag}a),
agreeing with the numerical results. The solution of the BdG equation also
yields the minimum energy gap (minigap) above the zero energy states. At
this gap and above there are other, finite-energy, states localized at the
end points of the wire. The minigap therefore provides the protection for
the Majorana states from finite temperature thermal effects. We see that the
minigap is of the same order of the pairing gap $\Delta \simeq 1$ meV, which
means that the Majorana fermion physics can be accessed at the
experimentally accessible temperatures $T<10$ K.

\emph{Effects of multiple confinement bands:}\textbf{\ }In a realistic
experiment, multiple confinement energy bands along the $y$ and $z$
directions need be taken into account \cite{Lutchyn,Multi} because they are
mixed with each other by the large spin-orbit coupling. Considering the
lowest $n$ confinement bands, the BdG Hamiltonian can be written as an $%
8n\times 8n$ matrix similar as Eq. (\ref{BdG}) with the matrix elements
replaced with the corresponding multiband forms. Specifically, $H_{1}\left(
x\right) $ is replaced with $H_{\text{n}}(x)=\int dydz(\phi _{1}^{\ast
}(y,z),...\phi _{n}^{\ast }(y,z))^{T}H_{L}(\phi _{1}(y,z),...\phi _{n}(y,z))$%
, where $\phi _{i}(y,z)$ is the wavefunction on the $i$-th band in the $yz$
plane. $\Upsilon $ is replaced with $\Upsilon _{n}=I_{n}\otimes \Upsilon $.
The Nambu spinor basis becomes $\hat{\Psi}(x)=(\hat{\psi}_{1}(x),\cdots ,%
\hat{\psi}_{n}(x),\Upsilon \hat{\psi}_{1}^{\dag }(x),\cdots ,\Upsilon \hat{%
\psi}_{n}^{\dag }(x))^{T}$ with $\hat{\psi}_{i}(x)=(\hat{\psi}_{\frac{3}{2}%
i}(x),\hat{\psi}_{\frac{1}{2}i}(x),\hat{\psi}_{-\frac{1}{2}i}(x),\hat{\psi}%
_{-\frac{3}{2}i}(x))^{T}$ as the hole annihilation operator on the $i$-th
band. We also assume there is no superconducting pairing between holes at
different confinement bands, therefore $\Delta _{Sn}=I_{n}\otimes \Delta
_{S} $. Following a similar procedure as that for the single band, we derive
the topological index $\mathcal{M}=$ sgn$[$Pf$\left\{ \Delta _{Sn}\Upsilon
_{n}\right\} $Pf$\left\{ \Upsilon _{n}\Delta _{Sn}+H_{n}^{T}\left( k\right)
\Upsilon _{n}\Delta _{Sn}^{-1}H_{n}\left( k\right) \right\} ]$ for the
multiband model.

Here we consider only the lowest two relevant energy bands ($l_{y}=1\ $and $%
2 $, $l_{z}=1$). In Fig. 3c, we plot the parameter regime for the TS state
when the multiple confinement induced bands are included. The parameter
regime for the TS state now has some quantitative difference from that in
the one-confinement-band case. However the basic conclusion remains the
same, that is, the Majorana fermions can be realized even with very small
Zeeman fields. In Fig. 3d, we plot the energies of the ground and first
excited states by solving the relevant BdG equations in the multiband model.
The figures confirm that the two methods -- topological index and numerical
solutions of BdG equations -- yield the same results. In the multiband
model, the minigap is slightly reduced, but still at the same order of $%
\Delta $.

In practice, $\Delta $ may depend on the material as well as the interface
between the superconductor and the nanowire. Although the parameter regime
for the emergence of Majorana fermions does not change much as a function of
$\Delta $, the minigap has a strong dependence on $\Delta $. In Fig. 2b, we
plot the minigap with respect to $\Delta $ in the multiband model and find
that $E_{g}\sim \Delta $ (similar as the electron-doped case \cite{Long-PRB}%
), instead of $\Delta ^{2}$ as in a regular $s$-wave or a chiral-$p$ wave
superconductor. Therefore the minigap is rather large, which ensures thermal
robustness of the Majorana fermions.

\emph{Conclusion:} The list of systems capable of supporting a non-Abelian
TS state now includes the filling factor $\nu =5/2$ FQH state, chiral-$p$
wave superconductors/superfluids, topological insulators, and electron-doped
semiconductors. To this list we have added a new system, a \emph{hole}-doped
nanowire, as a possible non-Abelian platform. Although the roster is
growing, ours is not an ordinary addition. As we have shown here in detail,
the requirements (carrier density, magnetic field, \textit{g-}factor,
\textit{etc.}) for the TS state in a hole-doped nanowire are \emph{already
accessible in experiments}. Thus this system can be a potential breakthrough
facilitating solid-state demonstration of Majorana fermions as well as
realization of TQC using a nanowire network.

\textbf{Acknowledgements:} We thank A. Akhmerov, F. Hassler, and M. Wimmer
for helpful discussion on the form of the superconducting pairing. This work
is supported by DARPA-MTO (FA9550-10-1-0497), DARPA-YFA (N66001-10-1-4025),
ARO (W911NF-09-1-0248), and NSF-PHY (1104546).


\begin{thebibliography}{99}
\bibitem{Bp} An anti-symmetric matrix $A=\left(
\begin{array}{cc}
R & Q \\
-Q^{T} & S%
\end{array}%
\right) $ can be diagonalized as $PAP^{T}=\left(
\begin{array}{cc}
R & 0 \\
0 & S+Q^{T}R^{-1}Q%
\end{array}%
\right) $ with $P=\left(
\begin{array}{cc}
I & 0 \\
Q^{T}R^{-1} & I%
\end{array}%
\right) $. Here $R$, $S$, $Q$ are matrices, $I$ is a unit matrix.
\end{thebibliography}
\end{document}